# Kagome lattice candidate URhSn: Measured Bragg diffraction patterns and theory for hexagonal and trigonal Sohncke phases

S. W. Lovesey[1, 2, 3] and D. D. Khalyavin[1]

[1] ISIS Facility, STFC, Didcot, Oxfordshire OX11 0QX, UK

[2] Diamond Light Source, Harwell Science and Innovation Campus, Didcot, Oxfordshire OX11 0DE, UK

[3] Department of Physics, Oxford University, Oxford OX1 3PU, UK

**ABSTRACT** Ternary intermetallics use a ZrNiAl-type chemical structure known as a kagome system. Complexity observed in the phase diagrams of these compounds is attributed to the geometric frustration. The 5f-electron compound URhSn exhibits two phase transitions at 16 K and 54 K. No super-lattice reflections in neutron powder diffraction patterns are detected in the intermediate range of temperatures (16 K - 54 K). Recent resonant x-ray diffraction patterns for uranium ions in the intermediate phase suggest a chiral structure and a trigonal Sohncke space group. It is firmly established that a simple hexagonal space group describes the parent structure at room temperature. A ferroic order parameter that includes hexadecapoles is shown to characterize the trigonal Sohncke phase. Resonant x-ray diffraction by this phase is fully investigated with our symmetry informed theory. It repairs symmetry in the original data analysis by using a grey group, and it lifts an unjustified spatial condition on uranium positions. The theory uses axial atomic multipoles that satisfy sum-rules established in the first instance for x-ray dichroic signals. A confrontation between observed and calculated diffraction patterns lends support to a trigonal Sohncke phase, and exposes grounds for additional experimental work on an enigmatic compound.

## I. INTRODUCTION

Properties of URhSn that make it of current interest include a kagome lattice, participation of atomic 5f electrons - rather than d electrons - and two successive phase transitions [1-6]. A ferromagnetic transition occurs at $T_c \approx 16$ K, and a second-order phase transition at $T_o \approx 54$ K. Neutron powder diffraction detected no super-lattice reflections between $T_c$ and $T_o$ [1]. On the basis of their resonant x-ray Bragg diffraction patterns, Tabata *et al*. propose a trigonal Sohncke phase for the temperature interval ($T_c$ - $T_o$) that descends in symmetry from a hexagonal phase (parent) [2]. In which case, atomic uranium quadrupoles and hexadecapoles are permitted in a URhSn order parameter [7]. Bragg diffraction can unveil hidden orders of electrons, so called because they do not respond to techniques that frequent standard laboratory set-ups [8]. By way of an example, atomic neptunium quadrupoles and hexadecapoles provide a successful analysis of resonant x-ray Bragg diffraction by $NpO_2$., cf. Section 5 in Ref. [9].

Continuing with the subject material URhSn, we simulate resonant x-ray Bragg diffraction patterns using a symmetry informed theory applied to the hexagonal parent and a trigonal Sohncke phase. Required parameters match those reported in a Letter by Tabata *et al*., e.g., a fundamental reflection vector [2]. In the range of temperatures between $T_c$ and $T_o$, additional Bragg spot intensity was detected superposed onto fundamental reflections. Upon cooling below $T_c$, reported as a ferromagnetic phase along the crystal c axis, substantial enhancements were observed, and interpreted as a simultaneous emergence of in-plane spin alongside the c axis ferromagnetic components [2]. Furthermore, observed behaviours were loosely interpreted as antiferro-quadrupole (AFQ) order, and a zero-propagation vector **k** = (0, 0, 0) which coexists with a ferromagnetic component below $T_c$. The resulting ground state structure breaks the mirror symmetry perpendicular to the kagome plane. In the actual experiment, Bragg spots intensities were enhanced by tuning the energy of the primary beam of x-rays to the M edge of uranium ions. An electric dipole -electric dipole (E1-E1) absorption event samples 5f valence states, namely, $M_{4,5}$ edges 3d → 5f. This electric dipole event is relatively weak compared to $L_{2,3}$ 2p → 3d in 3d transition metal compounds [16-18].

The trigonal phase of URhSn proposed by Tabata *et al.* is a chiral crystal structure, because its space group contains only proper operations [2,10]. The 65 Sohncke space groups include 11 enantiomorphous pairs that are usually set apart. In detail, the intermediate phase of URhSn is described by the trigonal Sohncke grey group P3211′ (No. 150.26, crystal class 3211′, $D_3$). The descent in symmetry from the hexagonal parent occurs through a loss of the mirror perpendicular to the c axis. Unit cells for the hexagonal parent (No. 189.222, $D_{3h}$) and trigonal Sohncke subgroup are depicted in Fig. 1 [2]. A scalar product of an anapole (Dirac dipole [5]) and an axial magnetic moment is the accepted hallmark of chirality of a material [11-14]. The monopole in question has been extensively discussed for an electric dipole - magnetic dipole (E1-M1) absorption event [§4 9, 11, 15]. Specifically, natural circular dichroism that detects enantiomorphic structure [11,12].

Unlike Tabata *et al.* [2] our x-ray diffraction amplitudes are not restricted to Wyckoff positions with cylindrical symmetry [19]. Time reversal (1′) in grey groups that define the hexagonal and trigonal Sohncke phases forbid magnetic (time odd) multipoles. Axial magnetic dipoles listed in Table I of Ref. [2] are not allowed, for example. Uranium ions satisfy oriented 3g position symmetry m2m1′ for ions in the hexagonal phase [20]. Spatial symmetry operations include mirrors $m_2' = \sigma_\theta \sigma_\pi 2_y$ with **y** = **b***, and $m_z' = \sigma_\theta \sigma_\pi 2_z$ with **z** = **c**. Our parity signature $\sigma_\pi$ = +1 (−1) for an axial (polar) multipole, and the time signature $\sigma_\theta$ = +1 (−1) for a non-magnetic (magnetic) multipole. Monoclinic position symmetry $2_x'$ is retained for U ions in the trigonal phase and the two mirror symmetries are lost. The AFQ order of quadrupoles mentioned by Tabata *et al.* [2] is a local arrangement of uranium quadrupoles (multipole rank K = 2) with (yz) spatial symmetry, depicted in Fig. 1. A ferroic order parameter is non-magnetic in a grey group, and the minimum rank of a multipole in the trigonal phase exceeds an octupole (K = 3).

## II. RESONANT X-RAY DIFFRACTION EXPERIMENT

Resonant x-ray Bragg diffraction amplitudes are delineated in Fig. 2. Atomic multipoles possess discrete symmetries in space and time. For an E1-E1 absorption event a multipole is represented by $\langle T^K_Q \rangle$ with (2K + 1) projections $-K \leq Q \leq K$. The multipole is invariant with respect to all symmetries in its Wyckoff position, in accord with Neumann's principle [21, 22]. Angular brackets ⟨...⟩ denote a time-average of the enclosed spherical tensor operator [23]. Real and imaginary parts of a multipole obey the sign convention $\langle T^K_Q \rangle = [\langle T^K_Q \rangle' + i\langle T^K_Q \rangle'']$, and a complex conjugate $\langle T^K_Q \rangle^* = (-1)^Q \langle T^K_{-Q} \rangle$ [23]. Multipoles for an E1-E1 absorption event are axial ($\sigma_\pi$ = +1) and posses a time signature $\sigma_\theta = (-1)^K$ [9]. Multipoles with odd K are forbidden in a grey group ($\sigma_\theta$ = +1). Wyckoff positions for U ions in URhSn are not centres of inversion symmetry and polar ($\sigma_\pi$ = −1) varieties of multipoles are permitted; Dirac multipoles are forbidden in grey groups [23, 24]. Scagnoli and Lovesey provide diffraction amplitudes for an electric dipole - electric quadrupole (E1-E2) absorption events that are not explored in any detail in the present study [25]. Higher-order electronic multipoles have been studied in a variety of compounds, including, monoclinic $LiCoPO_4$ and orthorhombic $LiNiPO_4$ [15], CuO [§4 9], $V_2O_3$ [26] and $\alpha$-$Fe_2O_3$ [27]. Experiments on URhSn are reported for a fundamental reflection vector **a*** = $(2\pi/a)$ (1, 1/√3, 0) where the cell edge $a$ ≈ 7.3580 Å [1]. In complete alliance with Tabata *et al.* [2], orthogonal local axis (x, y, z) mentioned in the caption to Fig. 2 coincide with standard hexagonal axes **a** = ($a$, 0, 0), **b*** and **c** = (0, 0, $c$) with $c$ ≈ 3.9826 Å [1]. The reciprocal lattice vector **b*** = [1, 2, 0], and **a** and **a*** enclose an angle = 30°.

Diffraction amplitudes are labelled by photon polarization, e.g., ($\pi'\sigma$) is an amplitude for primary x-ray polarization $\sigma$ normal to the plane of scattering and rotated to $\pi'$ within the plane, cf. Fig. 2. A resonant atomic process may dominate all other contributions to the x-ray scattering length should the photon energy E match a sharp atomic resonance with an energy $\Delta$ [2, 9, 16, 17, 18]. Assuming virtual intermediate states in the process are spherically symmetric, to a good approximation, the x-ray scattering length $\propto \{(\mu\eta)/(E - \Delta + i\,\Gamma/2)\}$in the region of the resonance, where $\Gamma$ is the total width of the resonance [23]. The numerator ($\mu\eta$) in the scattering length is the aforementioned amplitude for Bragg diffraction in the scattering channel with primary (secondary) polarization $\eta(\mu)$. Axial electronic multipoles $\langle T^K_Q\rangle$ used in E1-E1, and E2-E2, diffraction amplitudes comply with sum-rules established for dichroic signals [23, 28].

Our diffraction amplitudes allow for an azimuthal angle scan in which the illuminated crystal is rotated through an angle $\psi$ about the reflection vector. Fig. 3 includes experimental results reported by Tabata *et al.* [2]. Configurations 1 and 2 therein are replicated by azimuthal angles $\psi = 0°$ and $\psi = 90°$, respectively. The present study of resonant x-ray Bragg diffraction amounts to an application of the example treated by Scagnoli and Lovesey, Appendix B [25].

## III. HEXAGONAL PHASE $P\bar{6}2m1'$ (No. 189.222, BNS)

Amplitudes for an E1-E1 absorption event use electronic multipoles $\langle T^K_Q\rangle$ with ranks K = 0, 2 and projections Q = 0, ±2. Even K is imposed by the parity-even time signature $\sigma_\theta = (-1)^K$ and $\sigma_\theta$ = +1 in a grey group [9]. Even Q is imposed by the mirror symmetry $m_z' = \sigma_\theta \sigma_\pi 2_z$ in the uranium position symmetry. Bragg diffraction amplitudes for a reflection vector ($h$, 0, 0) reveal two uranium quadrupoles,

$$(\sigma'\sigma) = -A^0{}_0 + (1/(2\sqrt{2}))\,[3\cos(2\psi) + 1]\,A^2{}_0 - \sqrt{3}\sin^2(\psi)\,A^2{}_2, \qquad (1)$$

$$(\pi'\sigma) = -(1/2)\sin(\theta)\sin(2\psi)\,[\sqrt{(3/2)}\,A^2{}_0 + A^2{}_2] + i\cos(\theta)\sin(\psi)\,B^2{}_2,$$

$$(\pi'\pi) = -\cos(2\theta)\,A^0{}_0 + (1/\sqrt{2})\,[\sin^2(\theta)\,(3\cos^2(\psi) - 1) - 1]\,A^2{}_0$$

$$+ \sqrt{3}\,[1 - (\sin(\theta)\,\sin(\psi))^2]\,A^2_2.$$

Rotated and unrotated amplitudes are odd and even functions of $\psi$, respectively. Thomson scattering $A^0_0$ is excluded from $(\pi'\sigma)$ [25]. It appears in $(\sigma'\sigma)$ and $(\pi'\pi)$ for a lattice (nuclear) Bragg reflection $(h, 0, 0)$. Diffraction in the unrotated channel $(\sigma'\sigma)$ is independent of the Bragg angle $\theta$ depicted in Fig. 2 [25]. Amplitudes $(\sigma'\pi)$ and $(\pi'\sigma)$ are related by the sign of all $A^K_Q$ [25]. Referring to Appendix A, $A^K_Q$ and $B^K_Q$ are purely real and proportional to $\langle T^K_Q\rangle'$, e.g., $B^2_2(189) = [\sqrt{3}\,\langle T^2_{+2}\rangle'\,\sin(2\pi h x)]$, where the general coordinate $x \approx 0.5860$. Regarding angular anisotropy in allowed quadrupoles, they resemble $\langle T^2_0\rangle \propto (3z^2 - 1)$ and $\langle T^2_{+2}\rangle' \propto (x^2 - y^2)$. A notable feature of $(\pi'\sigma)$ is a 90° phase separation between $B^2_2$ and $A^K_Q$, and intensity is schematically $|(\pi'\sigma)|^2 \propto [A^2 + B^2]$. In addition, we shall see that the phase separation creates a chiral signature.

Intensity added to a Bragg spot by circular polarization in the primary x-ray beam = $P_2\Upsilon$ where a chiral signature [27, 29],

$$\Upsilon = \{(\sigma'\pi)^*(\sigma'\sigma) + (\pi'\pi)^*(\pi'\sigma)\}''. \tag{2}$$

The Stokes parameter $P_2$ (a purely real pseudoscalar) measures helicity in the x-ray beam. Evidently,

$$\Upsilon(189) = \cos(\theta)\,\sin(\psi)\,B^2_2\,[(\pi'\pi) - (\sigma'\sigma)]. \tag{3}$$

Intensity in the rotated channel of polarization $|(\pi'\sigma)|^2 = 0$ for Configuration 1 ($\psi = 0°$), likewise $\Upsilon = 0$. Referring to Fig. 3, a null value of $|(\pi'\sigma)|^2$ appears to disagree with an increase in intensity for temperatures below $T_o \approx 54$ K, cf. Fig. 3 [2]. Intensity observed in Configuration 2 ($\psi$ = 90°) can be attributed to $|(\pi'\sigma)|^2 = [\cos(\theta)\,B^2_2]^2$. Potentially significant intensities $|(\pi'\pi)|^2$ are permitted in the hexagonal phase for both experimental configurations and not much can be learned from available measurements. A null value of $|(\pi'\sigma)|^2$ below $T_o$ is overturned by reduced symmetry in the trigonal Sohncke phase.

## IV. TRIGONAL SOHNCKE P3211′ (No. 150.26, BNS)

Concerning bulk properties of uranium ions in the trigonal phase, a monopole $\mathbf{U}^0$ contributes natural circular dichroism to an E1-M1 absorption event [11]. It has another role, however, as the definition of chirality in a material [11, 14]. Operational forms include $\mathbf{U}^0 = (\mathbf{S} \cdot \mathbf{\Omega}_S)$, where $\mathbf{\Omega}_S = (\mathbf{S} \times \mathbf{R})$ is the spin anapole [5], and $\mathbf{S}$ and $\mathbf{R}$ are electronic spin and position variables, respectively. An alternative anapole uses orbital angular momentum $\mathbf{L}$, and $\mathbf{\Omega}_L = [\mathbf{L} \times \mathbf{R} - \mathbf{R} \times \mathbf{L}]$ [23]. Both forms of the monopole are explicitly time even (non-magnetic) and polar ($\mathbf{S}$ and $\mathbf{L}$ are axial). The amplitude for rotated polarization $(\pi'\sigma)$ in an E1-M1 absorption event includes $\langle\mathbf{U}^0\rangle$, where it is independent of the azimuthal angle understandably. The monopole is not present in $(\sigma'\sigma)$ and $(\pi'\pi)$ for E1-M1.

Inspection of the electronic structure factor Eq. (A2) shows that ferroic orders of multipoles are allowed. Evidently, for a null reflection vector it can be different from zero for integer (Q/3). Trigonal multipoles with projections Q = ±3 are allowed, and hexagonal multipoles are restricted to even Q. An E2-E2 absorption event uses hexadecapoles (K = 4) and it exposes $\langle T^4_{\pm 3}\rangle$ [25].

Returning to the parity even E1-E1 absorption event, projections Q are unrestricted by Wyckoff position symmetry in the trigonal Sohncke phase. Many details for the phase are relegated to Appendix A to avoid clutter here in the main text. In brief, trigonal amplitudes are a sum of contributions in Eq. (1) and contributions $A^2_1$ and $B^2_1$ created by the quadrupole $\langle T^2_{+1}\rangle''$. Explicit expressions for $A^K_Q$ and $B^K_Q$ in terms of E1-E1 axial multipoles are listed in Appendix A. The trigonal amplitude $(\pi'\sigma)$ can be expressed as,

$$(\pi'\sigma)(\text{No. } 150) = (\pi'\sigma)(\text{No. } 189) - i\sin(\theta)\cos(2\psi)\, A^2_1 - \cos(\theta)\cos(\psi)\, B^2_1. \qquad (4)$$

Here, $(\pi'\sigma)$(No. 189) is Eq. (1) evaluated with $A^K_Q$ and $B^K_Q$ in Appendix A for P3211′. Notably, contributions in Eq. (4) specific to the trigonal phase can be different from zero in Configuration 1. They answer the fault $(\pi'\sigma)$(No. 189) = 0 for $\psi = 0^o$. Moreover, $A^2_1$ forms the imaginary part of $(\pi'\sigma)$(No. 150). The multipole $B^2_1$ is absent in Configuration 2 ($\psi = 90^o$). Schmitt *et al.* review technical advantages in favour of measuring $|(\pi'\sigma)|^2$ instead of intensities in unrotated channels of polarization [30].

A result for the chiral signature $\Upsilon(150)$ for an E1-E1 is relatively simple for Configurations 1 and 2 because amplitudes $(\pi'\pi)$ and $(\sigma'\sigma)$ are purely real for $\psi = 0^o$ and $90^o$. From Eq. (2) we find,

$$\Upsilon(150) = \sin(\theta)\, \langle T^2_{+1}\rangle''\, \alpha''\, [(\pi'\pi) + (\sigma'\sigma)]. \qquad (5)$$

The factor $[(\pi'\pi) + (\sigma'\sigma)]$ is obtained from Eqs. (1) and (A6) evaluated for the appropriate azimuthal angle. Recall that $\Upsilon(189) = 0$ in Configuration 1. A chiral signature Eq. (2) calculated for Sohncke-type $Pb(TiO)Cu4(PO_4)_4$ does not include Thomson scattering [32]. The reason being that the reflection vector is not a lattice vector, unlike the present calculation where, in alliance with Tabata *et al.* [2], we use **a*** and $A^0_0$ in $(\pi'\pi)$ and $(\sigma'\sigma)$ can be different from zero.

## V. CONCLUSIONS

In summary, we have explored properties of non-magnetic hexagonal and trigonal Sohncke structures in view of the room-temperature parent and lower temperature intermediate phases of URhSn, respectively [2] (grey groups No. 189.222 and No. 150.26, BNS [20]). Uranium electronic octupoles and hexadecapoles appear in a ferroic order parameter. Hayami and Kusunose have prepared instructive coloured renditions of these and other multipoles [33]. Symmetry informed calculations of resonant x-ray Bragg diffraction patterns show no major shortcomings when confronted with measurements, albeit a sparse set of results, cf. Fig. 3 [2]. The mentioned hexadecapole belongs to the trigonal Sohncke phase and it will contribute to some diffraction patterns. Chiral signatures and polar multipoles have not been measured. They are predicted to expose differences between the hexagonal and trigonal phases.

Simulations of electronic structure - band structure calculations - are among other ways of achieving some of our results. They are a form of experiment in their own right. The accuracy of results depends on handling electronic correlations, among other things. By way of recent examples, density-functional theory as implemented in the ELK code, has been used to study Mn multipoles in the rutile antiferromagnet $MnF_2$ [34, 35]. The FDMNES code is specifically designed to simulate x-ray absorption and resonant scattering [36]. It has been used to estimate diffraction by $CoF_2$, a simple compound not unlike $MnF_2$ [37].

**ACKNOWLEDGEMENT** S. W. L. has benefitted from 27 years' instruction from Dr K. S. Knight (ISIS) on crystallography and Bragg diffraction culminating in insights to Sohncke lattices.

## APPENDIX A: Axial diffraction amplitudes

Hexagonal base vectors **a** = ($a$, 0, 0), **b** = ($a$/2) (−1, √3, 0), **c** = (0, 0, $c$). URhSn cell dimensions at 85 K; $a \approx 7.3580$ Å, $c \approx 3.9826$ Å [1].

Central to symmetry informed studies of diffraction is an electronic structure factor,

$$\Psi^K_Q = \sum \exp(i\, \boldsymbol{\kappa} \cdot \mathbf{d}) \langle T^K_Q \rangle_\mathbf{d}. \qquad \text{(A1)}$$

The sum is over positions **d** used by U ions in a unit cell, and the reflection vector $\boldsymbol{\kappa} = (h, k, l)$. The electronic structure factor includes information about the relevant Wyckoff positions found in the Bilbao table MWYCKPOS [20]. Wyckoff positions are related by operations listed in the table MGENPOS [20]. Taken together, the two tables provide all information required to complete Eq. (A1). Parity-even ($\sigma_\pi = +1$) multipoles $\langle T^K_Q \rangle$ with even K apply to E1-E1 and electric quadrupole - electric quadrupole (E2-E2) events.

Parent $P\bar{6}2m1'$, No. 189.222 [20]; reflection vector ($h$, 0, 0) with $\mathbf{a}^* = (2\pi/a)\, (1, 1/\surd3, 0)$ and $\sin(\theta) = [h\, (2\lambda/3a)]$. We find,

$$\Psi^K_Q = \langle T^K_Q \rangle\, [\alpha + f_Q + (f_Q\, \alpha)^*], \qquad \text{(A2)}$$

with $\alpha = \exp(i2\pi h x)$, and $f_Q = \exp(i2\pi Q/3)$. The result Eq. (A2) is different for hexagonal and trigonal phases because they are endowed with different uranium multipoles. This is readily apparent in bulk properties with $\alpha = +1$ ($h = 0$), and Q = ±6 for the hexagonal phase and Q = ±3 in the trigonal phase. The trigonal (ferroic) order parameter $\langle T^4_{\pm3} \rangle$ contributes to diffraction enhanced by an E2-E2 event.

In the footsteps of Scagnoli and Lovesey, the electronic structure factor $\Psi^K_Q$ is partitioned in even and odd functions of Q, with a notation $A^K_{-Q} = A^K_Q$, $B^K_{-Q} = -B^K_Q$ [25]. The authors treat the case in hand as a worked example. Following suit,

$$[A^K_Q + B^K_Q] = \exp(iQ\chi)\, \Psi^K_Q, \qquad \text{(A3)}$$

and $\chi = 150^\circ$ since **a** and **a*** enclose an angle = 30°. Thomson scattering at lattice (nuclear) reflections is fixed by $A^0_0(189) = \Psi^0_0(189)$ and,

$$A^2_2(189) = [\langle T^2_{+2}\rangle' (\alpha' - 1)],\ B^2_2(189) = [\surd 3\ \langle T^2_{+2}\rangle' \alpha''], \qquad (A4)$$

with $\alpha' = \cos(2\pi h x) = 0.050$ for $h = 3$ [2]. Diffraction amplitudes for the hexagonal phase are recorded in Eq. (1).

Trigonal Sohncke P3211′, No. 150.26.

The Wyckoff position symmetry monoclinic $2_x$ admits odd Q. E1-E1 diffraction amplitudes are the same as for the hexagonal phase with addition of a quadrupole $\langle T^2_{+1}\rangle''$, and different coefficients in $A^K_Q$ and $B^K_Q$. Immediate confrontation of theory and the data analysis reported by Tabata *et al*. [2] is concluded with trigonal factors for E1-E1 absorption, namely,

$$A^2_0 = \langle T^2_0\rangle' (2\alpha' + 1),\ A^2_1 = \surd 3\ \langle T^2_{+1}\rangle'' \alpha'',\ B^2_1 = -\langle T^2_{+1}\rangle'' (\alpha' - 1),$$

$$A^2_2 = \langle T^2_{+2}\rangle' (\alpha' - 1),\ B^2_2 = -i\surd 3\ \langle T^2_{+2}\rangle' \alpha''. \qquad (A5)$$

Common structure in hexagonal and trigonal amplitudes in regard to the Bragg angle θ and azimuthal angle ψ allow shorthand versions of diffraction amplitudes for the trigonal phase, namely,

$$(\sigma'\sigma)(\text{No. }150) = (\sigma'\sigma)(\text{No. }189) - i\surd 3 \sin(2\psi)\ A^2_1, \qquad (A6)$$

$$(\pi'\pi)(\text{No. }150) = (\pi'\pi)(\text{No. }189) - i\surd 3 \sin^2(\theta) \sin(2\psi)\ A^2_1.$$

The result for $(\pi'\sigma)$(No. 150) is in Eq. (2). The imaginary part of $(\pi'\sigma)$(No. 150) is provided by a quadrupole $A^2_1 \propto (yz)$. No such contributions occur in the hexagonal phase. These multipoles are responsible for a chiral signature $\Upsilon(150)$.

*Parent* hexagonal space group $P\bar{6}2m$ (No. 189). Wyckoff positions;

U 3g (0.5860, 0.0000, 0.5000), symmetry m2m ($2_x$, $m_z$, $m_{120}$)
Rh 3f (0.2470, 0.0000, 0.0000)
Sn1 1a (0.0000, 0.0000, 0.0000)
Sn2 2d (0.3333, 0.6667, 0.5000)

*Trigonal subgroup* Sohncke P321 (No. 150) keeps the lattice vector and origin of the parent structure. Wyckoff positions;
U 3f (0.5860, 0.0000, 0.50000), symmetry .2. ($2_x$)
Rh 3e (0.2470, 0.0000, 0.0000)
Sn1 1a (0.0000, 0.0000, 0.0000)
Sn2 2d (0.3333, 0.6667, 0.5000)

## APPENDIX B: Polar diffraction amplitudes

Polar multipoles $\langle \mathbf{U}^K\rangle$ ($\sigma_\pi = -1$, $\sigma_\theta = +1$) mentioned in Section IV contribute to diffraction by the hexagonal and trigonal Sohncke phases. We consider the latter. Positions 3f in P3211′ used by uranium ions are not centres of inversion symmetry, and Wyckoff symmetry alone requires $\langle U^K_Q\rangle = [(-1)^{K+Q} \langle U^K_Q\rangle^*]$. By dint of the "totalitarian principle" of symmetry, attributed to Murray Gell-Mann [24], polar multipoles are compulsory. Multipole ranks in an E1-E2 event are K = 1, 2, 3, and all contributions appear in Appendix D in Ref. [25]. E1-E1 and E1-E2 resonances occur at different photon energies. Octupole contributions include the ferroic order parameter $\langle U^3_{\pm3}\rangle'$.

Confining ourselves to dipoles and some quadrupoles, Q = 0, ±1, ±2, and the electronic structure factor (A2) and definition (A3) yield,

$$A^1_0 = 0,\ A^1_1 = i\langle U^1_{+1}\rangle' (\alpha' - 1),\ B^1_1 = -i\sqrt{3}\ \langle U^1_{+1}\rangle' \alpha'',\ A^2_0 = \langle U^2_0\rangle' (2\alpha' + 1), \quad \text{(B1)}$$

$$A^2_1 = \sqrt{3}\ \langle U^2_{+1}\rangle'' \alpha'',\ B^2_1 = -\ \langle U^2_{+1}\rangle'' (\alpha' - 1),\ A^2_2 = \langle U^2_{+2}\rangle' (\alpha' - 1),\ B^2_2 = -i\sqrt{3}\ \langle U^2_{+2}\rangle' \alpha''.$$

At the level to which we work for the moment, the corresponding amplitude in the rotated channel of polarization,

$$(\pi'\sigma)(\text{No. } 150) \approx \sin(2\theta)\ \sin(\psi)\ [A^1_1 + (i2/3)\ \sqrt{5}\ B^2_1] \quad \text{(B2)}$$

$$+\ i\sqrt{(5/6)}\ [\cos^2(\psi)\ \{3 - 5\sin^2(\theta)\} - 2 + 3\sin^2(\theta)]\ A^2_0 +\ \{\sqrt{5}/(6)\}\ \sin(2\psi)\ [5\cos(2\theta) + 1]\ A^2_1.$$

The contribution $A^2_1 \propto \langle U^2_{+1}\rangle''$ in Eq. (B2) differs from the other three by a factor $i$. It does not contribute to Configurations 1 and 2, cf. Fig. 3. However, multipoles $A^1_1$ and $B^2_1$ might make a significant difference between the two configurations.

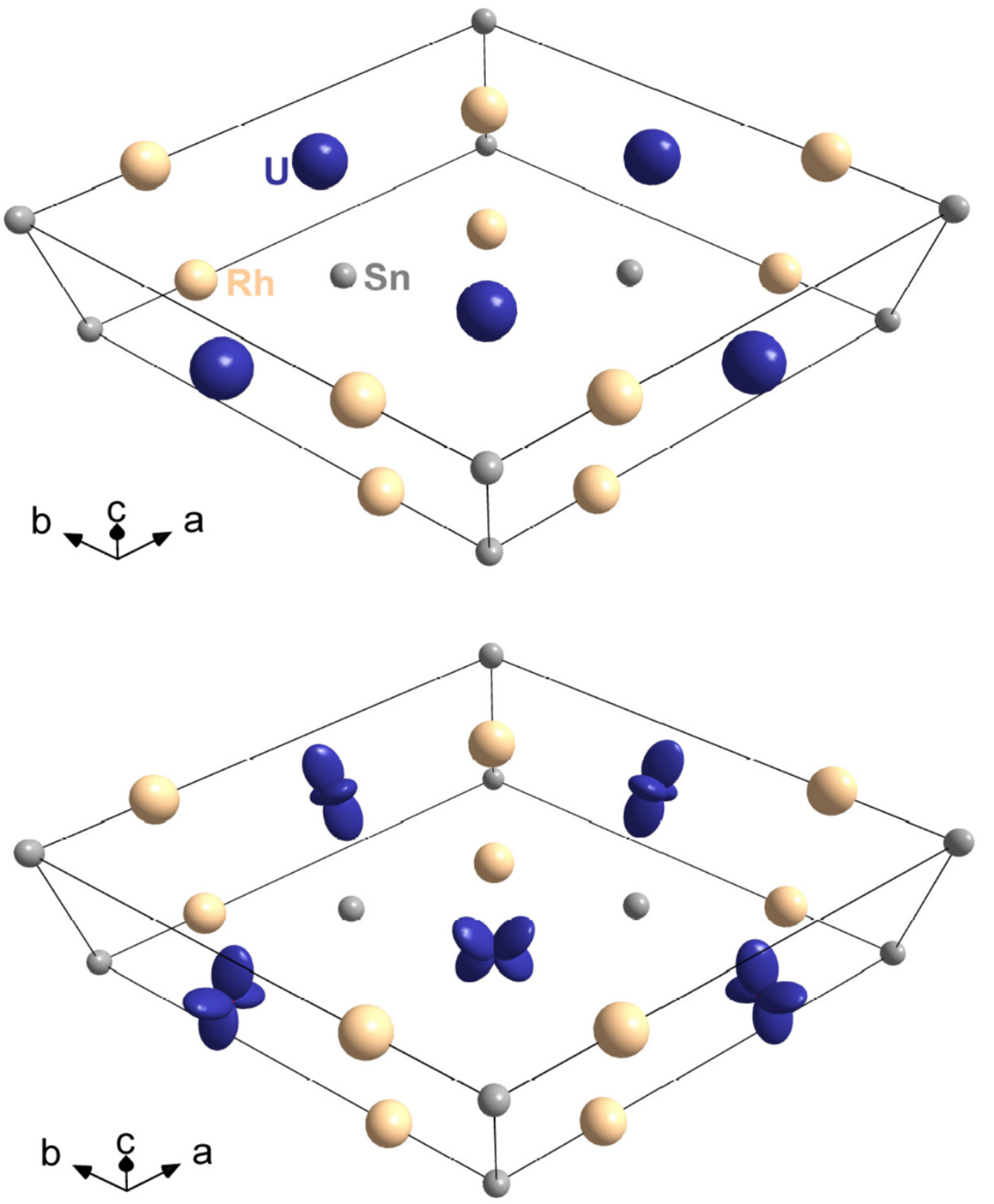

**FIG. 1.** URhSn: trigonal Sohncke phase, P3211′ (BNS [20]). Depiction of ions in a cell (upper panel) and spatial anisotropy of uranium quadrupoles $\langle \mathbf{T}^2 \rangle$ (bottom panel). Specifically, the quadrupole $A^2{}_1 \propto (yz)$ in Eq. (A6) [Appendix A]. Uranium ions in Wyckoff positions 3f are not centres of inversion symmetry and host polar multipoles $\langle \mathbf{U}^K \rangle$ (not shown, Appendix B) with discrete symmetries $\sigma_\pi = -1$ and $\sigma_\theta = +1$.

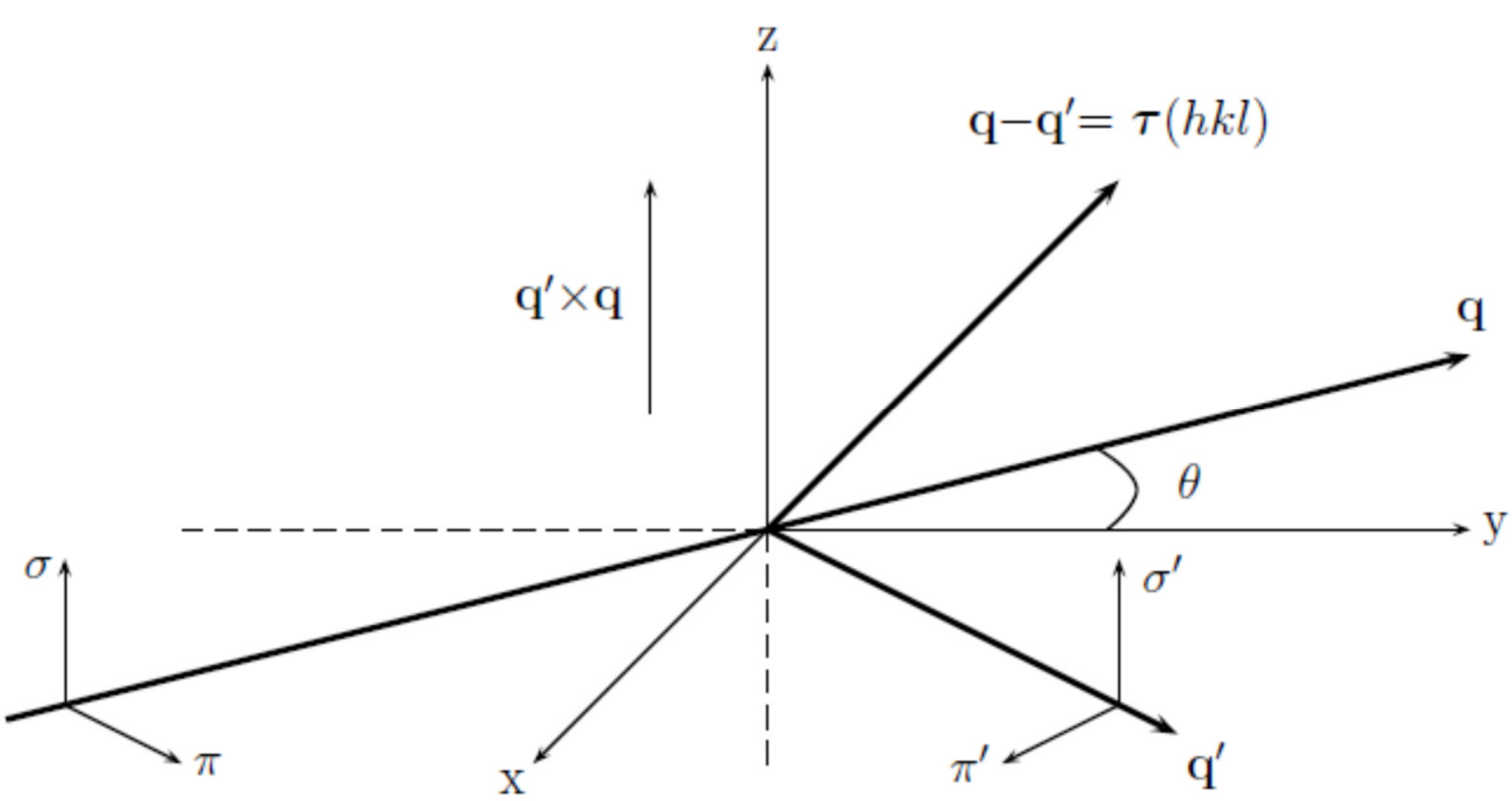

**FIG. 2.** Primary (σ, π) and secondary (σ′, π′) states of x-ray polarization; $\boldsymbol{\sigma} = \boldsymbol{\sigma}' = (0, 0, 1)$, $\boldsymbol{\pi}$

= (cos(θ), sin(θ), 0), **π**′ = (cos(θ), − sin(θ), 0). Corresponding wavevectors **q** and **q**′ subtend an angle 2θ, and θ is the Bragg angle. The depicted Cartesian (x, y, z) coincide with local axes (**a**, **b***, **c**) in the nominal setting of the crystal with an azimuthal angle ψ = 0; **b*** = (**a**, 2**b**, 0).

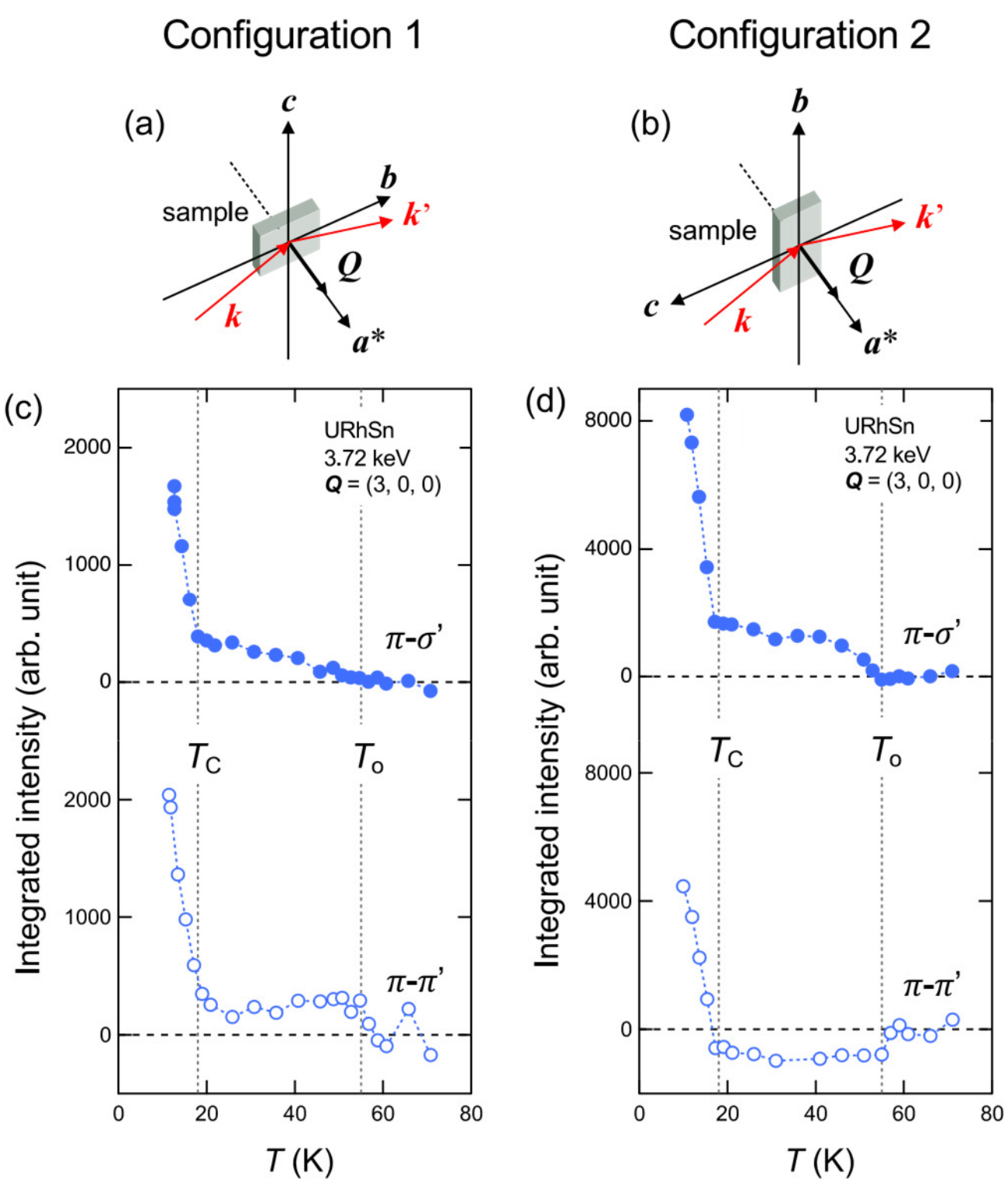

**FIG. 3.** Resonant x-ray diffraction patterns for URhSn as functions of temperature in unrotated and rotated polarization channels (labelled (π′π) and (π′σ) in the text to be consistent with Scagnoli and Lovesey [25]). Two phase transitions at 16 K and 54 K. Reflection vector ($h$**a***, 0, 0) is labelled **Q** in the figure and **κ** in Eq.(A1), Miller index $h$ = 3. Configuration 1 (ψ = 0°) and Configuration 2 (ψ = 90°). Reproduced from Ref. [2].